\documentstyle[sprocl,epsf]{article}

\begin{document} 

\title{CRITICAL AND CHAOTIC BEHAVIORS OF
QUARKS}\author{ RUDOLPH C. HWA }

\address{Institute of Theoretical Science and Department of Physics\\
University of Oregon, Eugene, OR 97403-5203, USA\\E-mail:
hwa@oregon.uoregon.edu}

\maketitle\abstracts{The critical behavior of quarks undergoing phase
transition to hadrons is considered in the framework of the Ising
model.  It is found that spatial fluctuations do not alter the $F$-scaling
result obtained earlier in the Ginzberg-Landau formalism.  For the
study of chaos a new measure must be found for the quark-gluon
system, for which the usual description appropriate for classical
nonlinear systems is inapplicable.  It is shown that the entropy indices
$\mu_q$ which characterize the fluctuations of spatial patterns, are as
effective as the Lyapunov exponent in measuring chaos.  When applied
to the QCD parton showers, it is found that the quark jets are more
chaotic than gluon jets.  The analysis is highly appropriate for a
quantitative treatment of the erratic fluctuations of multiparticle
production in leptonic, hadronic and nuclear collision processes.  It
represents a step beyond the traditional intermittency analysis.}

\section{Introduction}
	
The topics to be discussed here concern the nature of fluctuations of
hadronic observables in high-energy collisions.  The first is on
quark-hadron phase phase transition, a subject that has been
investigated from many approaches.  Our emphasis is on the scaling
behavior of what can be observed experimentally.  Our earlier study of
the subject has been on the basis of a mean-field theory.  We now
improve upon that by incorporating spatial fluctuations.
	
The second topic is on the possibility of chaotic behaviors of a
quark-gluon system.  Since the number of degrees of freedom of such a
system is not conserved, the traditional method of following the
distance between two nearby trajectories of a classical system cannot
be applied.  Thus a major part of our effort in that problem is to
identify an appropriate measure of chaos and to show that it is as good
as the usual one when applied to the classical problems.  With the new
measure it is then possible to investigate the question of chaos for QCD
parton showers.  The significance of this line of investigation is in
uncovering the fluctuating properties of multiparticle production that
have hitherto been averaged over.  The experimental determination of
the key measures that quantify those properties will then present a
challenge to theoretical models, which may not have been sufficiently
accurate (at least in soft interaction) to reproduce them.
	
The two topics will be discussed separately below, with more space
given to the latter, since it involves newer and more unfamiliar
concepts.

\section{Critical Behavior}

In a series of papers \cite{hn,for} the consequences of quark-hadron
phase transition in terms of hadronic observables have been studied in
the framework of the Ginzburg-Landau formalism.  It is found that the
normalized factorial moments $F_q$ do not exhibit intermittency (i.e.,
the power-law behavior
$M^{\varphi _q}$), but they do satisfy
$F$-scaling
	\begin{eqnarray}
 F_q \propto F^{\beta _q}_2  \quad ,  
\label{1}
\end{eqnarray} where
\begin{eqnarray}
 \beta _q = (q - 1)^{\nu}\, ,\quad \nu = 1.304  
\quad . 
\label{2}
\end{eqnarray} 
The value of the scaling exponent  $\nu$, which bears
no relationship to any of the usual critical exponents, is a definitive
characteristic of second-order phase transition, in which the
temperature is not a measurable quantity, and the observables are
multiplicities of the quanta produced.  Heavy-ion collision experiments
have shown the validity of (\ref{1}), but currently the value of
$\nu$ is not 1.304, signifying the absence of quark-hadron phase
transition.  But in a laser experiment at the threshold of lasing,
where the physics is known to be that of a second-order phase
transition, it has been shown that (\ref{2}) is valid to a high degree of
accuracy.\cite{yea}

The Ginzburg-Landau theory is a mean-field theory, so the spatial
properties of the system are assumed to be smooth. To improve on that
description, it is necessary to take into account the fluctuating
character of hadronization.  That has been done recently in two
different directions.  One is done by simulation on a 2D lattice using the
Ising model for hadronization.\cite{cgh}  The other is also done on the
lattice, but surface fluctuations (perpendicular to the lattice) are
introduced with the thermal effects being constrained by the color
confinement potential.\cite{zch}   We discuss them in turn.

\subsection{Spatial fluctuations simulated on the Ising lattice}

In the Ginzburg-Landau (GL) approach to the problem of phase
transition we relate the order parameter $\phi(z)$ to the hadron
density by $\rho(z) = \left| \phi(z)\right|^2$.  For a bin of area
$\delta^2$ we assume that $\rho(z)$ is constant in the bin so the GL
free energy is simplified to the form $F[\phi] = \delta^2 \left( a\left|
\phi\right|^2 + b\left|\phi\right|^4 \right)$.  The multiplicity
distribution $P_n$ of hadrons for $T < T_c$ is then given
by\cite{hn,for}
\begin{eqnarray}
 P_n = Z^{-1}\int {\cal{D}} \phi \, \, {1 \over n!}\left( \delta ^2 \left|\phi
\right|^2 \right)^n e ^{-\delta^2 \left|\phi
\right|^2 - F[\phi]}
\label{3}
\end{eqnarray}
where $Z = \int {\cal{D}} \phi e ^{- F[\phi]}$.  To improve on
this,\cite{cgh} we allow $\phi(z)$ to vary within a bin by adapting the
GL theory of ferromagnetism, \cite{bea} using the Ising model for
$\phi(z)$
\begin{equation}
\phi(z) = A^{-1/2}_{\varepsilon} \sum_{j \epsilon A_{\varepsilon}} s_j
\quad ,
\label{4}
\end{equation}
 where $s_j$ is the spin $\pm$1 at site $j$, and $z$ is now the location
of the center of a cell of size $A_{\varepsilon}$, which is $<\delta^2$,
but large enough to contain several sites.  The hadron multiplicity at
the
$i$th cell is then 
\begin{equation}
n_i = \lambda \left| \sum_{j \in
A_{\varepsilon}(i)} s_j \right|^2 \theta \left(\sum_j s_j \right)
\label{5}
\end{equation}
where $\lambda$ is a scale factor that relates the quark
density of the plasma at $T_c$ to the lattice site density in the Ising
model.  The bin multiplicity is $n_{\delta} = \sum^{N_{\delta}}_{i = 1}
n_i$, where $N_{\delta} = \left( \delta/\varepsilon\right)^2$.
	
By using the Wolff algorithm \cite{bea} to simulate the bin
configurations in the Ising model, we can calculate $P_n(\delta)$ and
therefore $F_q(M)$, where $M = (L/\delta)^2$, $L$ being the lattice
size.  We have found that the strict scaling behavior, $F_q \propto
M^{\varphi _q}$, is valid only at $T = T_c$.  In fact, $T_c$ can be
determined by varying $T$ until that behavior is manifested.  However,
for $T \leq T_c$ the more general $F$-scaling behavior (1) is valid for a
range of $T$.  Furthermore, (2) is also valid, but the value of $\nu$
depends on $T$.

Our result is that $\nu \simeq 1.0$ at $T_c$ but $\nu$ becomes bigger
at $T < T_c$.  For a range of $T < T_c$, the values of $\nu$ are between
1.0 and 1.6 so that the average value of $\nu$ is about 1.3.  This is
a very satisfying solution to a dilemma posed by an earlier analytical
result, where Satz \cite{hs} found $\varphi _q = (q - 1)/8$ at $T =
T_c$ in the 2D Ising model.  It means that $\beta _q = \varphi
_q/\varphi _2 = q-1$, so that $\nu = 1.0$.  Now, we see that the GL
mean-field result of $\nu = 1.304$ is an average of $\nu(T)$ over a
range of $T < T_c$, without invalidating the analytical result at
precisely $T = T_c$.

\subsection{Surface fluctuations}

There is another way to consider the spatial fluctuations of
hadronization sites inside a bin.  Imagining the 2D surface to be a
membrane that can have displacements normal to a flat reference
surface, then an outward protrusion from the plasma interior can be
identified as a bump where hadronization is more likely to occur
than a dent, an inward indentation.  The two competing mechanisms
that control the nature of the surface fluctuations are (a) the
thermal fluctuations, and (b) the confining potential on the partons
that prefers no deformation of the flat membrane.  For the latter we
can consider an additional term to the free energy, $F_s$, representing
an increase due to spatial displacement.  Specifically, we take it to be
$F_s = C\sum_{<ij>} \left| z_i - z_j\right|$, where $C$ is a parameter
proportional to the confinement strength, and $z_i$ is the vertical
displacement from a flat surface at site $i$.  $C$ contains the
thermal effect represented by $\left(k_B T \right)^{-1}$.  The
competition between the two mechanisms is then introduced by the
Boltzmann factor exp$(-F_s)$.

In our calculation \cite{zch} we use the GL formalism to describe
the mean result, which is to be convoluted with the fluctuation
result that is computed on a lattice by Monte Carlo simulation using
the exp$(-F_s)$ factor to determine the probability of nonzero $z_i$
displacements.  Consideration of cells in bins is again necessary, as in
the Ising problem described above.  Omitting the details we
summarize here that $F$-scaling as in (1) is still valid; moreover, the
slopes $\beta _q$ are nearly independent of $C$ in the range $0 < C
<2$, which includes the values that has the maximum effect due to
surface fluctuations.  The formula in (2) is also still valid, and the
value of $\nu$ is found to be 1.306 $\pm$ 0.035.

Our conclusion is then that the GL result \cite{hn,for} for the scaling
exponent $\nu$ for quark-hadron phase transition is unaffected, when
spatial fluctuations due to surface displacements are taken into
account.

\section{Chaotic Behavior}

We now come to fluctuations of a different nature.  For phase
transitions, because of the competition between the ordered
(collective) and disordered (thermal) motions of the constituents at
the critical point, there can be large fluctuations over extended
domains.  But for jets of hadrons produced in hard processes or for
multiparticle production at low $p_T$ in hadronic collisions, we do not
have dense thermal systems of partons (unless it is a heavy-ion
collision at high energy) and phase transition is not the sort of physics
that one is concerned with.    There are, nevertheless, fluctuations in
the event multiplicity and in the phase-space distribution of the
particles produced.  Are those fluctuations unpredictable?  In what
way can they be recognized as chaotic behavior?

In order to introduce a quantitative measure of unpredictability we
shall proceed in stages.  First, we shall discuss the notion of erraticity,
which concerns the fluctuations of the spatial patterns of the final
states in momentum space.  Then we shall discuss entropy indices,
followed by their applications to both the classical nonlinear problems
and the QCD parton showers.  The consideration involves an area of
phenomenology that has hitherto been unexplored.  

\subsection{Erraticity}

Intermittency refers to the scaling behavior of the multiplicity
fluctuations in bins of size $\delta$.\cite{bp}  In determining the
normalized factorial moments by averaging over all events at a fixed
bin and then averaging over all bins, information on the spatial
patterns from event to event is lost.  Erraticity analysis is an attempt to
capture that information.

Let $F_q$ denote a specific quantification of the spatial pattern in the
phase space of a final state.  It can be the horizontal moments, the
correlation integrals, or the result of wavelet analysis.  After many
events one obtains a distribution $P(F_q)$ of $F_q$.  Now, let us define
the normalized (vertical) moments
\cite{zch2}
	\begin{eqnarray} 
 C_{p,q} = \left< F_q^p\right>/\left<  F_q\right>^p   \quad ,
\label{6}
\end{eqnarray}
where the averages are done with $P(F_q)$ as the
probability distribution.  The fluctuations of $F_q$ become more simple
to categorize and more interesting to study, if
$C_{p,q}$ exhibits scaling behavior, i.e., 
 	\begin{eqnarray} C_{p,q} \propto M^{\psi_q(p)} \quad ,
\label{7}
\end{eqnarray} 
where $M$ is the number of bins in a fixed space.  For
convenience, we refer to this behavior as erraticity.\cite{rch2}  It is a
natural extension of the notion of intermittency.  The value of $p$ can
be any positive real number.  It should not be negative because $F_q$
may vanish for some events, so $P(F_q = 0) \not= 0$ for some $q$.  For
$0<p<1$, it is the $F_q < 1$ region that is probed by $\psi _q(p)$, while
for $p > 1$ it is the spiky events with high $F_q$ that $\psi _q(p)$
describes.  In practice it is not necessary to consider $p$ greater than 3.

Quantities similar to $C_{p,q}$ have been considered before.  In
statistical physics the random energy model \cite{ds} for spin glass has
been investigated with the consideration of the
quantity $\left<Z^p(\beta) \right>$, where $Z(\beta)$ is the partition
function $\sum_{\omega} \mbox{exp} \left[ -\beta E (\omega)\right]$,
$\beta$ being the inverse temperature.  It led Brax and
Peschanski \cite{bp2} to study in the $\alpha$ model the quantity
$\left<Z^p_q \right>$ where $Z_q = \sum_m (\rho_m/\sum_m
\rho_m)^q$,  $\rho _m$ being the density of particles in the {\it
m}th bin.  Since $q$ plays the role of $\beta$ in $Z(\beta)$, the
possibility of ``non-thermal'' phase transition is considered in Ref.\ 12. 
The same quantity
$\left<Z^p_q \right>$ has similarly been studied in Ref.\ 13 also in the
framework of the $\alpha$ model.  The emphasis of erraticity analysis
is not on the theoretical possibility of a phase transition, but on
determining the scaling behavior of $C_{p,q}$ from data (experimental
and model simulation) and on extracting the erraticity indices
$\psi_q(p)$.  It is reasonable to suggest that $\psi_q(p)$ provides a
stringent test of the reality of any model on multiparticle production.

The study of erraticity can be applied to many problems in physics,
chemistry and beyond.  Fluctuation in spatial patterns is a ubiquitous
phenomenon.  The Ising model, for instance, has many types of patterns
in the lattice space depending on the temperature.  One can even
determine the critical temperature for a finite lattice in any dimension
by studying the behavior of
$\psi_q(p)$.\cite{zch3}

It is also possible to determine the erraticity spectrum $e(\alpha)$
analogous to the multifractal spectrum $f(\alpha)$.  The definition for
$e(\alpha)$, for a fixed $q$, is \cite{rch2}   
  \begin{eqnarray}  e(\alpha) = p \alpha - \psi (p) \,  ,  \quad \quad
\alpha = d \psi (p)/dp
\quad  .
\label{8}
\end{eqnarray} At $p = 1$, $e(\alpha) =
\alpha$, since $\psi(1) = 0$.  That value of the spectrum has a
particular significance, as we shall discuss next.  The function
$e(\alpha)$ exhibits that value explicitly, unlike
$\psi(p)$.

\subsection{Entropy indices}

The unpredictability of the outcome of a dynamical process should be
related to some quantity like the entropy that measures the ignorance
about the system.  We do not want to follow the time evolution of the
system, which is not observable.  Our entropy must refer to an
ensemble of final states.  To that end we use
$F_q$ as in Sec.\ 3.1, to describe the spatial pattern of an event, and
$P(F_q)$ to describe the distribution of $F_q$ after
$\cal{N}$ events.  The average of any function
$f(F_q)$ can also be written as 
  \begin{eqnarray}
 \int dF_q\,  P(F_q) f(F_q) = {\cal N}^{-1}
\sum^{\cal N}_{e=1} f(F_q^e)
  \quad ,
\label{9}
  \end{eqnarray}  
where $F^e_q$ is the $F_q$ for the $e$th event.  Introduce now
  \begin{eqnarray}
 P_e = F_q^e/\sum^{\cal N}_{e=1}F_q^e \quad ,
\label{10}
  \end{eqnarray} in terms of which we can define an entropy in the
``event space''
  \begin{eqnarray}
 S = - \sum_e  P_e \mbox{ln} P_e\quad ,
\label{11}
  \end{eqnarray} 
(which should probably be called ``eventropy'', since it is not the
usual entropy.)  To calculate
$S$ it is more convenient to introduce the moments $H_p$ such that
\begin{eqnarray} S = - \left. {d \over dp}
\mbox{ln} H_p \right|_{p=1} \quad ,
\quad \quad H_p =
\sum_e \left(P_e\right)^p \quad .
\label{12}
  \end{eqnarray} 
From (\ref{9}) we have
\begin{eqnarray} H_p =  {\cal N} \int dF_q P(F_q) \left[{F_q \over {\cal
N} \int dF_q P(F_q)F_q}\right]^p = {{\cal N}\left<F^p_q\right>
\over \left<{\cal N}F_q\right>^p} = {\cal N}^{1-p}C_{p,q}
\quad , 
\label{13}
  \end{eqnarray} If the system exhibits erraticity, it then follows from
(\ref{7}) that
\begin{eqnarray} S = \mbox{ln} \left({\cal N} M ^{-\mu_q} \right)
\quad , 
\label{14}
  \end{eqnarray} where
\begin{eqnarray}
\mu _q = \left.{d \over dp}
\psi_q(p)\right|_{p=1}  \quad . 
\label{15}
  \end{eqnarray} 
We refer to $\mu_q$ as the entropy indices.\cite{zch2}

If $\mu _q = 0$, then $S = \mbox{ln} {\cal N}$, which is large.  One may
think that it describes a system that is chaotic.  Actually, the opposite
is true.  It corresponds to $P_e = 1/{\cal N}$ in (\ref{11}), i.\ e., every
event has the same value of $F_q$, according to (\ref{10}).  That is
hardly what one would expect of a chaotic system, since the spatial
patterns are the same for every event.  $S$ is large because in the
event space $F_q$ is evenly distributed over the entire space.  $S$
would be small if $F^e_q \not= 0$ is restricted to only a few events;
that would mean large fluctuations in $F^e_q$.  Thus finite,
nonvanishing positive values of $\mu _q$ corresponds to wide
$P(F_q)$, which in turn means unpredictable spatial pattern from event
to event.  In short, $\mu _q$ is a measure of unpredictability.

An alternative way of calculating $\mu _q$ is to circumvent $S$ and work directly with $C_{p,q}$ by expressing it as 
\begin{eqnarray}
C_{p,q} = \left< \Phi ^p _q\right> \quad , 
 \quad \quad \Phi  _q =
{F_q \over \left<F_q
\right>} 
  \quad . 
\label{16}
  \end{eqnarray}
Then we have
\begin{eqnarray}
\left.{d \over dp}C_{p,q}\right|_{p=1} = \left< \Phi  _q \; \mbox{ln}
\; \Phi  _q
\right> 
  \quad . 
\label{17}
  \end{eqnarray}
On the other hand, if $C_{p,q}$ has the scaling behavior (\ref{7}), then
we also have  
\begin{eqnarray}
\left.{d \over dp}C_{p,q}\right|_{p=1} = \mu _q \; \mbox{ln}
\; M 
\label{18}
  \end{eqnarray}
in the scaling range of $M$.  Consequently, we obtain
\begin{eqnarray}
\mu _q = {\partial \over \partial \mbox{ln} M } \left< \Phi  _q \;
\mbox{ln}
\; \Phi  _q
\right> 
  \quad , 
\label{19}
  \end{eqnarray}
which is to be determined in the region where $\sigma _q \equiv \left<
\Phi  _q \; \mbox{ln} \; \Phi  _q \right> $ exhibits a linear dependence
on $\mbox{ln} M$.

\subsection{Chaoticity}

We now consider the question whether multiparticle production
processes are chaotic.  Let us first recall the properties of chaotic
behavior in classical nonlinear dynamics.\cite{hgs}  Since in such
systems trajectories in space-time exist, the distance function $d(t)$
between two trajectories can be defined.  A system is chaotic if $d(t)
\sim e^{\lambda t}$, $\lambda > 0$, no matter how small $d(0) =
\epsilon$ may be, for $\epsilon > 0$.  For classical Yang-Mills dynamics
the Lyapunov exponent $\lambda$ has been shown to be positive by
lattice calculation.\cite{bm}

For quantized Yang-Mills fields the problem is vastly more
complicated.  In addition to the ambiguity associated with quantum
chaos in the realm of first quantization, we now have also the problems
of nonconservation of the number of degrees of freedom and of the
lack of a meaningful definition of trajectory.  The absence of an
unambiguous notion of time in production processes further results in
the nonexistence of the Lyapunov exponent for the problem.

Our first task must be the search for a measure of chaos appropriate
to our problem.  Since nonperturbative QCD is too difficult to
implement, we focus on parton showers in pQCD, i.\ e., quark and
gluon jets in a tree-diagram approximation of QCD branching processes. 
In place of classical trajectories whose initial points are all in a small
neighborhood of one another, we consider branching processes all
starting from exactly the same virtuality
$Q^2$ and let quantum fluctuations take them to different final states. 
Since we have no experimental access to the process of branching, and
theoretically the degradation of vituality does not have an unique
association with the temporal evolution, our measure of chaos must be
focused on what we can observe, {\it viz}, the characteristics of the
final states.  It is in that respect that the subject of erraticity and
entropy indices becomes relevant.

There are two issues here.  One is to find a measure of chaos, and the
other concerns QCD branching processes.  We discuss them separately.

\subsection*{3.3.1  A new measure of chaos}

Our proposal is to use $\mu _q$ as an alternative measure of chaos in
problems where only the spatial patterns can be observed.  Whether
that approach agrees with the conventional approach can be
determined only by applying the measure to known chaotic systems, of
which there are many with simple nonlinear dynamics.\cite{hgs}  They
are classical systems with well-defined trajectories, and the positivity
of the maximum Lyapunov exponent $\lambda$ is a sufficient criterion
for chaos.  Since our measure is concerned with spatial patterns rather
than temporal evolution, it is necessary to construct a spatial pattern
for every trajectory.  That is achieved by considering a set of points
corresponding to the positions of a trajectory at a discrete collection of
times, separated by finite intervals apart.  Clearly, two nearby
trajectories for $\lambda < 0$ would generate two very similar
patterns, while two chaotic trajectories for $\lambda > 0$ would have
very dissimilar patterns.  We can generate $\cal{N}$ events by
considering
$\cal{N}$ trajectories all starting from a small neighborhood of an
initial point.  Studying the
$P(F_q)$ distribution of those events and then comparing the resultant
$\mu_q$ to $\lambda$ constitute our procedure to check the
effectiveness of $\mu_q$ as a measure of chaos.

We have applied this procedure to the logistic map and the Lorenz
attractor, the details of which are omitted here.  They are nonlinear
systems that become chaotic when their control parameters $r$ exceed
certain critical values.\cite{hgs}  We present here only the result for the
logistic map.\cite{zch3}  In Fig.\ 1 the dashed line shows the value of
$\lambda$ as a function of $r$.  For $r > 3.57$, $\lambda$ becomes
positive, although there are short intervals where $\lambda$ drops
below zero.  The solid line indicates the value of
$\mu_2$ (normalized by a specific factor), calculated at discrete points
in the same range of
$r$.  Evidently
$\mu _2$ coincides very well with $\lambda$ except that it cannot be
negative but vanishes when $\lambda \leq 0$.  The multiplicative
factor used for $\mu _2$ in the plot depends on the number of spatial
points taken to determine the $P(F_q)$ distribution, and has no
particular significance.  What is of great significance is that we now
have an alternative measure of chaos.  The positivity of $\lambda$ is
coordinated with the positivity of $\mu _2$ (and all other $\mu _q$),
so that the fluctuations of spatial patterns can equally serve as a means
for revealing chaotic behaviors.

There are many complex systems possessing complex patterns.  The
determination of their entropy indices may reveal certain features that
are not otherwise recognizable.  For some nonlinear systems that are
known theoretically to be chaotic, the experimental verification of
$\lambda$ may be difficult, since the precise adjustment of the initial
condition may not always be possible.  Our method of studying the
spatial patterns may therefore offer a feasible alternative.

\begin{figure}
\centerline{\epsfbox{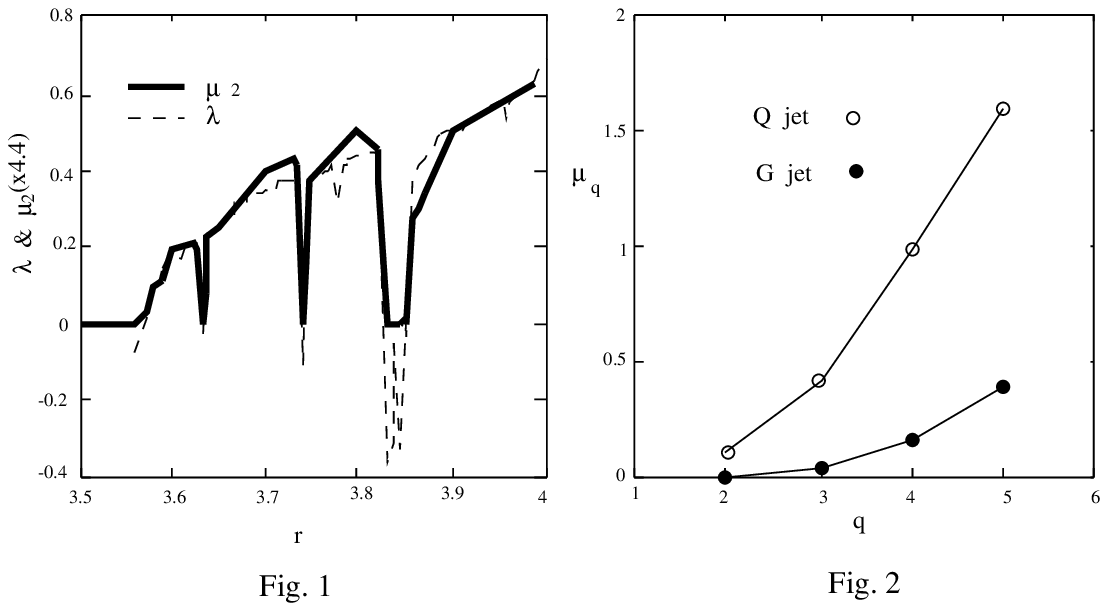}}
\end{figure}

\subsection*{3.3.2 Chaos in QCD parton showers}

Since perturbative QCD is well developed theoretically, and the final
states of quark and gluon jets are well measured experimentally, we
propose that the erraticity of hard processes be used as a new arena
for comprehensive tests and that the entropy indices of parton showers
be determined to reveal the chaoticity of QCD.

Since hadronization is irrelevant to the question of chaos in pQCD, we
have developed an efficient Monte Carlo generator of quark and gluon
jets with the usual Sudakov form factor and splitting functions
incorporated in the code.\cite{zch2}  The initial virtuality of a quark (Q
jet) or a gluon (G jet) is fixed at a $Q^2$ value for all events, and all
partons evolve and branch until the virtuality of a parton reaches $\leq
Q^2_0$, where branching is terminated for that parton.  From the
spatial pattern of each event in the cumulative variable \cite{bg} (in
terms of which the inclusive distribution is flat), we calculate the
horizontal normalized factorial moments $F_q$, then the distribution
$P(F_q)$, the vertical moments $C_{p,q}$, the erraticity indices
$\psi_q(p)$, and finally the entropy indices
$\mu_q$.  The results of the calculation for
$Q/Q_0 = 10^3$ are shown in Fig. 2.  The positivity of $\mu _q$ for
both Q and G jets indicates that the QCD branching processes are
chaotic.  Moreover, the Q jets are more chaotic than the G jets.  The
reason is that quark jets have fewer partons produced, so the
multiplicity fluctuations relative to the mean is larger.

We have also considered the fixed coupling problem, since by varying
$\alpha _s$ as a control parameter one can study the behavior of the
system at the onset of chaos.  It turns out that there is no threshold of
chaos.  When
$\alpha _s$ is small, the multiplicity is so small that $\mu _q$ is large
for both Q and G jets.  Thus chaoticity is an inherent property of QCD
dynamics and cannot be tuned out by decreasing $\alpha _s$, the only
tunable parameter that we can associate directly with the strength of
the nonlinear term.

Using the positivity of $\mu _q$ as a criterion for chaos, we have not
only found that the non-Abelian, nonlinear dynamics of QCD is chaotic,
but also come to realize that any quantum system in which the number
of quanta of interest is not conserved is likely to have positive $\mu
_q$.  Thus while chaotic behaviors are spectacular and remarkable
phenomena in classical dynamics, their occurrences seem to be
common and generic in quantum systems.  Perhaps the notion of chaos
should not be more emphasized in multiparticle production than the
realization that the unpredictability of the final states can be
quantified.  To calculate the entropy indices for all collision processes,
soft as well as hard, that can agree with the experimental data would be
a theoretical challenge.  Our results obtained should not be taken
quantitatively for comparison with experiments, since the MC code has
not been tuned to check the other features of parton showers. 
However, they have been effective in demonstrating the procedure of
determining $\mu _q$ and in elucidating the question of chaos in QCD.
 
\subsection{Hadronic and nuclear collisions}

For hadronic collisions we expect more fluctuations than in $e^+e^-$
annihilation, since there is impact-parameter variation from event to
event.  Even at fixed impact parameter there can be various
cut-Pomerons, each of which can have large variations in the final
states corresponding to many possible ways that the branching
processes can proceed, at least according to the geometrical branching
model (GBM) for soft processes.\cite{ch}  That does not even include
the consideration of hard subprocesses at high energies that would
further increase the degree of fluctuations.  Gianini has found in his
preliminary analysis of the old Fermilab data that $\mu _2$ is in excess
of 0.4.\cite{gg}  Cao is currently upgrading ECCO,\cite{ph} an event
generator that is based on GBM, by taking into account resonance
production.  The preliminary result on the entropy indices is that they
are also very large.

In heavy-ion collisions the fluctuations in impact parameter are
usually controlled by making cuts in $E_T$.  Because of the high
multiplicity per event not much has been observed in the
intermittency behavior.  More may be revealed in the erraticity
analysis.  I venture to speculate here what the results on
$\mu _q$ would be in the four possible scenarios:  with and without a
hadron-gas phase, and with and without a quark-to-hadron phase
transition.  If there is a hadron gas phase before hadrons are emitted
so that there is thermalization in the final state, then the emitted
hadrons are randomized.  Random spatial patterns of many particles
can differ significantly from event to event, so I would not expect
$\mu _q$ to be small whether or not there is a phase transition.  If
there is no hadron gas phase that randomizes the hadronic final states,
then one should expect to see the dynamical effects of the production
processes in the erraticity analysis.  If there is a phase transition, the
produced hadrons should exhibit long-range correlations that result in
clusters of all sizes (assuming second-order phase transition and
Kadanoff scaling).  In that case we should expect larger values of $\mu
_q$ than in the case of no phase transition. 
  
\section{Conclusion}

Significant progress has been made recently on the subject of
fluctuations in multiparticle production.  On critical behavior
considerations of fluctuations beyond the result of mean-field theory
have not altered the scaling exponent
$\nu$.  For noncritical phenomena new measures of fluctuations from
event to event are proposed.  Experimental determination of
$\nu$,
$\psi_q(p)$ and $\mu_q$ is urgently needed.  For $\nu$ it is necessary
to consider high $q$ moments.  For $\psi_q(p)$ and $\mu _q$ low
$q$ values like 2 and 3 are enough and $p$ in the range 0.5 to 2 is
sufficient.  It is highly likely that many existing event simulators may
not be able to reproduce the experimental data on those quantities. 
Significant change of
$\mu_q$ due to quark-gluon plasma formation would be very exciting.

\section*{Acknowledgments} I have benefitted from discussions with
R.\ Peschanski. The research underlying the contents of
this talk was done in collaboration with Z.\ Cao.  It was supported, in
part, by the U.S. Department of Energy under Grant No.
DE-FG03-96ER40972.

\end{document}